\documentclass[a4paper,11pt]{article}
\usepackage{typearea}\typearea{12}
\usepackage{ifpdf}
\ifpdf
  \usepackage{graphicx}
\else
  \usepackage[dvipdfmx]{graphicx}
\fi
\usepackage{epsfig,amsmath,amsfonts,amssymb}
\ifpdf
  \usepackage{hyperref}
\else
  \usepackage[dvipdfmx]{hyperref}
\fi
\hypersetup{hidelinks,
  pdftitle={Trees that can be grown in too many ways: Bouch's construction and a bound for every size},
  pdfauthor={Hal Tasaki}}
\newtheorem{theorem}{Theorem}[section]
\newtheorem{lemma}[theorem]{Lemma}

\newtheorem{proposition}[theorem]{Proposition}

\makeatletter
\@addtoreset{equation}{section}
\makeatother

\makeatletter
\long\def\@makecaption#1#2{{\small
\advance\leftskip1cm
\advance\rightskip1cm
\vskip\abovecaptionskip
\sbox\@tempboxa{#1: #2}%
\ifdim \wd\@tempboxa >\hsize
 #1: #2\par
\else
\global \@minipagefalse
\hb@xt@\hsize{\hfil\box\@tempboxa\hfil}%
\fi
\vskip\belowcaptionskip}}
\makeatother
\def\eq#1\en{\begin{equation}#1\end{equation}}  
\def\eqa#1\ena{\begin{align}#1\end{align}}
\def\eqg#1\eng{\begin{gather}#1\end{gather}}
\newcommand{\lb}[1]{\label{e:#1}}
\newcommand{\rlb}[1]{\eqref{e:#1}} 
\newcommand{\nl}{\notag\\}

\newcommand{\sumtwo}[2]%
{\mathop{\sum_{#1}}_{#2}}
\newcommand{\sumthree}[3]%
{\mathop{\mathop{\sum_{#1}}_{#2}}_{#3}}
\newcommand{\sumfour}[4]%
{\mathop{\mathop{\mathop{\sum_{#1}}_{#2}}_{#3}}_{#4}} 
\newcommand{\prodtwo}[2]%
{\mathop{\prod_{#1}}_{#2}}
\newcommand{\mintwo}[2]%
{\mathop{\min_{#1}}_{#2}}
\newcommand{\maxtwo}[2]%
{\mathop{\max_{#1}}_{#2}}
\newcommand{\maxthree}[3]%
{\mathop{\mathop{\max_{#1}}_{#2}}_{#3}}
\newcommand{\limtwo}[2]%
{\mathop{\lim_{#1}}_{#2}}
\newcommand{\suptwo}[2]%
{\mathop{\sup_{#1}}_{#2}}
\newcommand{\supthree}[3]%
{\mathop{\mathop{\sup_{#1}}_{#2}}_{#3}}
\newcommand{\supfour}[4]%
{\mathop{\mathop{\mathop{\sup_{#1}}_{#2}}_{#3}}_{#4}} 
\newcommand{\inftwo}[2]%
{\mathop{\inf_{#1}}_{#2}}
\newcommand{\infthree}[3]%
{\mathop{\mathop{\inf_{#1}}_{#2}}_{#3}}
\newcommand{\inffour}[4]%
{\mathop{\mathop{\mathop{\inf_{#1}}_{#2}}_{#3}}_{#4}} 
\newcommand{\bbZ}{\mathbb{Z}}
\newcommand{\qedm}{\rule{1.5mm}{3mm}}
\usepackage{color}
\definecolor{fluorescentpink}{rgb}{1.0, 0.08, 0.58}
\definecolor{forestgreen}{rgb}{0.13, 0.55, 0.13}
\newcommand{\CB}{C_{\mathrm B}}

\begin{document}

\noindent
{\Large\bf
Trees that can be grown in ``too many'' ways:\\
Bouch's construction and a bound for every size}

\renewcommand{\thefootnote}{\fnsymbol{footnote}}
\medskip\noindent
{\small Hal Tasaki}\footnote{%
Department of Physics, Gakushuin University, Mejiro, Toshima-ku,
Tokyo 171-8588, Japan.
}
\renewcommand{\thefootnote}{\arabic{footnote}}
\setcounter{footnote}{0}

\vspace{-2mm}

\begin{quotation}
\small\noindent
We show that, for every positive integer $L$, there is a rooted tree on the square lattice that can be grown in at least $L!/C^L$ distinct ways, where $L$ denotes the number of bonds and $C>1$ is a constant independent of $L$.
Bouch's hierarchical construction \cite{Bouch2015} gives such trees along an unbounded sequence of sizes.
We carefully review his construction and combine it with an elementary extraction lemma to obtain the bound for every size.
(As discussed in Section~IV.A of \cite{ParkerCaoAvdoshkinScaffidiAltman2019}, Appendix~A.3 of \cite{ShiraishiTasaki2024}, and the video \cite{TasakiVideo2025}, the existence of such trees has an implication for operator growth and imaginary-time evolution in quantum spin systems in two or higher dimensions.)

\smallskip\noindent
I wrote the first version of this note as a self-contained exposition of Bouch's construction.
The stronger lower bound for every tree size presented here was proved by ChatGPT on the basis of that version (see the end of Section~\ref{S:introduction}).
I wish to make clear that I made no creative contribution to this beautiful and, I hope, useful theorem, whose proof is due to Bouch and ChatGPT.
\end{quotation}

%%%%%%%%%%%%%%%%%%
\tableofcontents
%%%%%%%%%%%%%%%%%%

\section{Introduction and the main theorem}\label{S:introduction}
Let $T$ be a rooted tree with $L$ bonds on the square lattice $\bbZ^2$.
More precisely, $T$ is a set of $L$ bonds (a bond or, equivalently, an edge is an unordered pair $\{u,v\}$ of sites $u,v\in\bbZ^2$ such that $|u-v|=1$) that is connected and contains no loops.
The root is a specified site that is attached to one of the bonds in $T$.
Throughout the note, the size of a tree means its number of bonds, not its number of sites.

\begin{figure}
\centerline{\epsfig{file=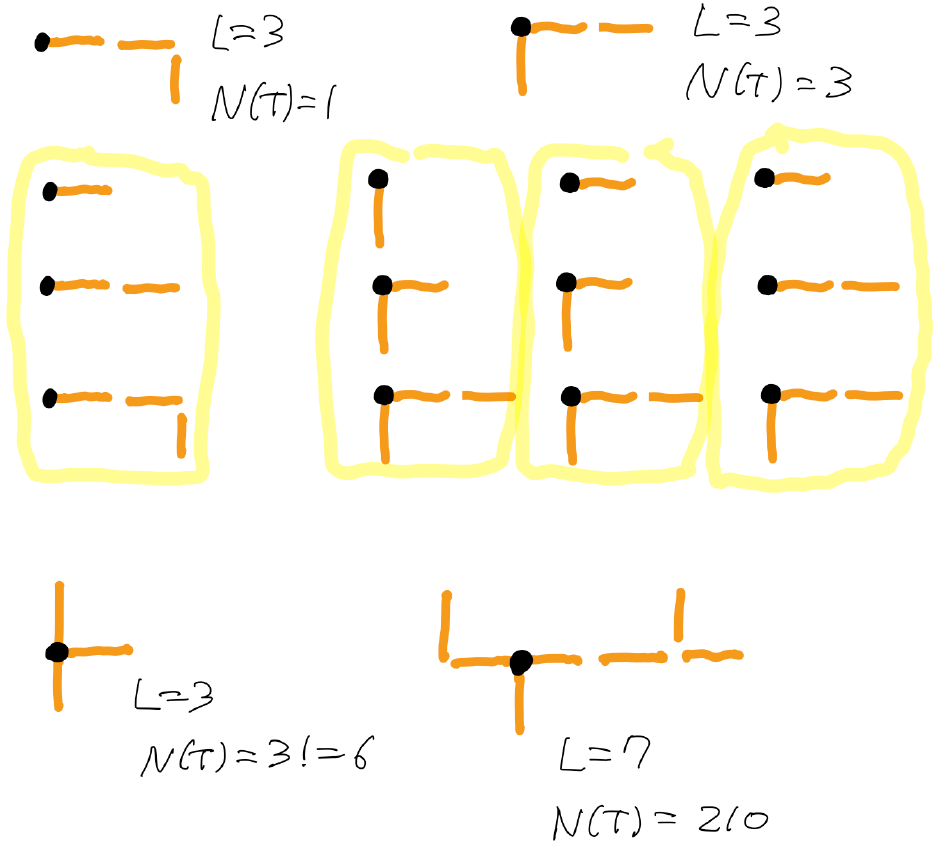,width=7truecm}}
\caption[dummy]{
Examples of rooted trees and the number of ways to construct them.
See Figure~\ref{f:weights} for the computation of $N(T)=210$ for the final example.
}
\label{f:trees}
\end{figure}

Suppose that one grows a given tree $T$ by starting from the root and adding bonds one by one.
We demand that, at every step, the root together with the bonds already added forms a connected graph.
Clearly, a tree that consists of a single path with the root at one end can be grown in a unique way.
But a general tree with branches can be grown in multiple ways.
See Figure~\ref{f:trees} for typical examples.

Let $N(T)$ denote the number of distinct ways that a given rooted tree $T$ can be grown.
It obviously holds that
\eq
1\le N(T)\le L!.
\lb{apriori}
\en
We are interested in how large $N(T)$ can be.
For example, the comb-like tree
\par
\begin{figure}[ht]
\centerline{\epsfig{file=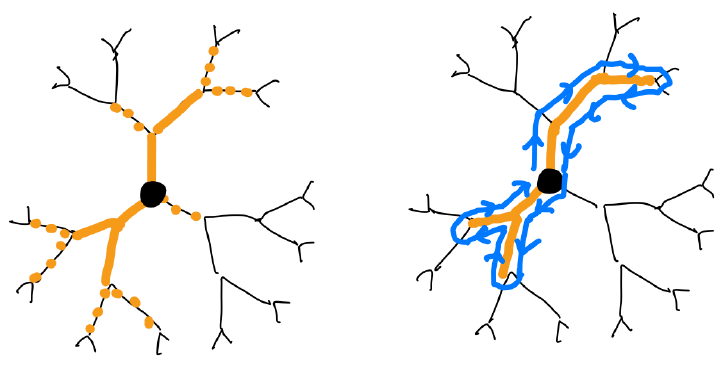,width=6truecm}}
\caption[dummy]{
For simplicity, we consider the Bethe lattice with coordination number three.
We first note that the number of ways to grow a tree with $L$ bonds from a fixed root is $3\times4\times5\times\cdots\times(L+2)$.
To see this, note that when the growing tree has $n$ bonds, one can add the next bond to one of the $n+3$ positions at the boundary.
The left figure depicts the case with $n=5$, where there are eight candidates shown by dotted lines.

We next observe that the number of distinct trees with $L$ bonds, on the other hand, is upper bounded by $3^{2L}$.
This is because any tree with $L$ bonds can be realized as the track of a walk with $2L$ steps that starts at the root.
See the right figure.

We thus conclude that there exists a tree that can be grown in at least $3\times\ldots\times(L+2)/9^L>L!/9^L$ ways.
Note that the proof is non-constructive.
}
\label{f:Bethe}
\end{figure}

\nopagebreak[4]
\centerline{\epsfig{file=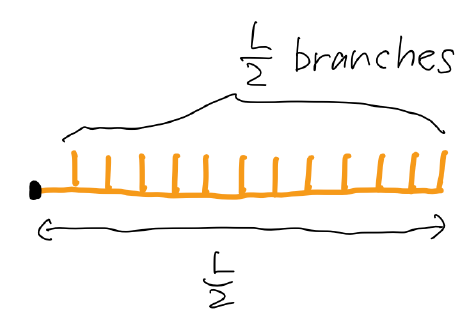,width=4truecm}}
\par\noindent
with even $L$ can be obtained by first growing the horizontal segment with $\frac{L}{2}$ bonds (in a unique manner) and then adding $\frac{L}{2}$ vertical branches in an arbitrary order.
This proves the lower bound $N(T)\ge(\frac{L}{2})!$.\footnote{
Lemma~\ref{L:Kupi} allows us to compute the exact number $N(T)=(L-1)!!$.
}
It is obvious that one can devise examples in which $N(T)$ grows much faster in $L$.
A nontrivial question is whether there exist trees that almost saturate the a priori upper bound in \rlb{apriori}, i.e., whether there is a constant $C>1$ such that, for any $L$, there is a tree $T$ with $L$ bonds satisfying
\eq
N(T)\ge\frac{L!}{C^L}.
\lb{main}
\en
If we consider the same problem on the Bethe lattice (or, equivalently, the infinite Cayley tree) instead of $\bbZ^2$, then the claim can easily be justified by a counting argument.
See Figure~\ref{f:Bethe}.

The following theorem gives an affirmative answer for trees on $\bbZ^2$ (and hence for those on $\bbZ^d$ with $d\ge2$).

\begin{theorem}\label{T:main}
There is a constant $C>1$ such that, for every positive integer $L$, there exists a rooted tree on $\bbZ^2$ with exactly $L$ bonds that satisfies \rlb{main}.
\end{theorem}

Bouch proved in 2015 that the bound \rlb{main} holds for an infinite set of positive integers $L$ \cite{Bouch2015}.
His proof is based on an ingenious hierarchical construction of a sequence of trees.
(He acknowledges J\'ozsef Beck for a suggestion.)
The additional observation in the present note is that Bouch's result already implies Theorem~\ref{T:main}.
An elementary extraction lemma allows us to pass from arbitrarily large trees to a tree of any prescribed size.
If $\CB>1$ is a constant in Bouch's bound, we may take $C=\CB^2$ in \rlb{main}.
No information about the gaps between the sizes in Bouch's sequence is needed.

In Section~\ref{S:extraction}, we introduce the weight of a rooted tree, prove the extraction lemma, and deduce Theorem~\ref{T:main} from Bouch's result.
In Section~\ref{S:Bouch}, we give a complete presentation of Bouch's construction and proof.
By introducing new symbols, reorganizing the derivation, and simplifying some estimates, we try to make Bouch's work accessible to a wide range of readers.

\medskip\noindent
{\em Note:}\/ Bouch \cite{Bouch2015} was motivated by a problem in quantum many-body physics.
His construction was a key to his proof of the existence of a singularity in the imaginary-time evolution of quantum spin systems in two or higher dimensions.

While working on the material in Appendix~A.3 of \cite{ShiraishiTasaki2024}, I realized that the existence of trees with the property \rlb{main} has an implication for the growth of Lanczos coefficients in quantum spin systems.
I tried to prove the existence of such trees, only to realize that it was a nontrivial problem.
Then I noticed that Bouch's paper was cited in many papers on the subject of operator growth, including the seminal work \cite{ParkerCaoAvdoshkinScaffidiAltman2019}.
The relation to imaginary-time evolution and operator growth is also discussed in my video \cite{TasakiVideo2025}.

I worked out the first version of this note mainly for myself to be sure that Bouch had the desired example.
I then decided that the note was worth making public because Bouch's result has important implications for quantum many-body physics and, moreover, is interesting by itself.

\medskip\noindent
{\em Note on the revision and the use of ChatGPT:}\/ The first version of this note reviewed Bouch's construction and left open whether \rlb{main} holds for every $L$.
I supplied that version to ChatGPT (OpenAI, GPT-6 Astra Pro, September 2026) and asked it to prove the all-sizes statement.
Without a suggestion from me about how to proceed, ChatGPT obtained the extraction lemma in Section~\ref{S:lemma} and used it to deduce Theorem~\ref{T:main} from Bouch's result.
It also prepared the draft of this revised note, including the reorganization of the proof.\footnote{
I then made some minor edits (including this footnote) and drew Figure~\ref{f:TU} because I genuinely enjoyed reading the proof provided by ChatGPT.
}
My contributions are the exposition, drawings, and editorial revisions; I do not claim a contribution to the additional proof.
Bouch's construction, reviewed in Section~\ref{S:Bouch}, remains the essential mathematical input.

\section{From arbitrarily large trees to every size}\label{S:extraction}
\subsection{The weight of a rooted tree}
\begin{figure}[ht]
\centerline{\epsfig{file=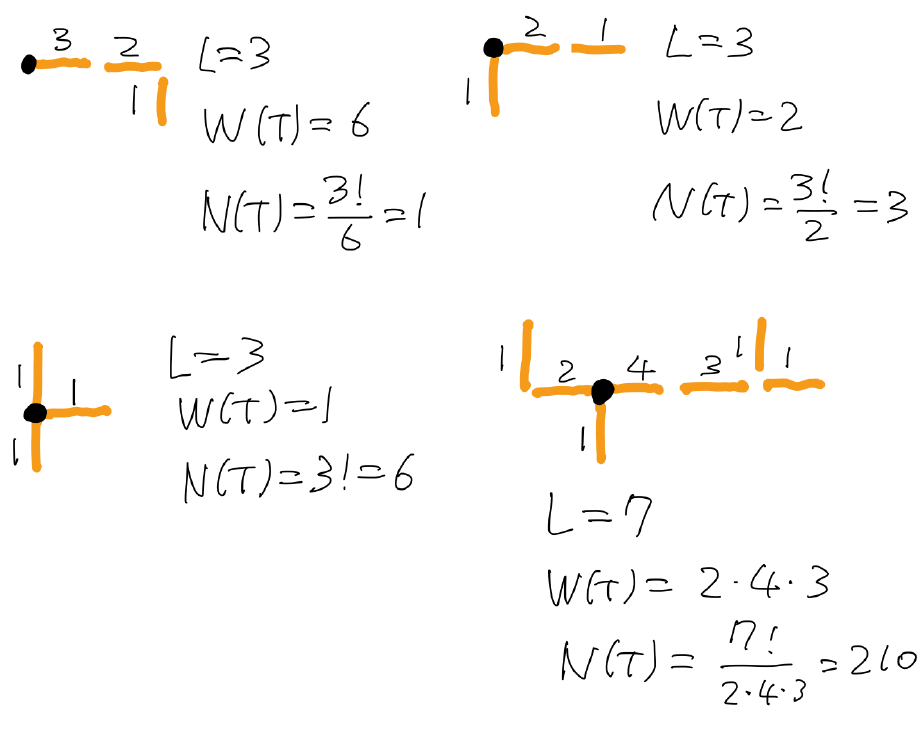,width=8truecm}}
\caption[dummy]{
The weights of bonds and trees in some examples.
The number of ways to grow a tree, $N(T)$, is computed by using the formula \rlb{NW}.
}
\label{f:weights}
\end{figure}

Let $T$ be a finite rooted tree.
For each bond $b\in T$ we define its weight $w_T(b)\in\{1,2,\ldots\}$ such that $w_T(b)-1$ represents the number of bonds in $T$ ``downstream'' of $b$.
(We understand that the ``flow'' originates at the root.)
We write $w(b)$ instead of $w_T(b)$ when the tree is clear from the context.
See Figure~\ref{f:weights}.

We then define the weight of the tree $T$ as
\eq
W(T)=\prod_{b\in T}w_T(b).
\lb{W}
\en
The following lemma (which is attributed in \cite{Bouch2015} to Elizabeth Kupin) states that the weight $W(T)$ determines the desired quantity, $N(T)$.
See Lemma~6.2 of \cite{Bouch2015}.
We include the elementary proof for completeness.

\begin{lemma}\label{L:Kupi}
The number of distinct ways that a given rooted tree $T$ with $L$ bonds can be grown is
\eq
N(T)=\frac{L!}{W(T)}.
\lb{NW}
\en
\end{lemma}

\noindent
{\bf Proof.}
For the induction, we also allow a tree consisting of a single root and no bonds, and set its weight and number of growth histories equal to one.
Suppose that the root of $T$ has $d$ branches with $n_1,\ldots,n_d$ bonds, where the first bond of each branch is included in $n_i$.
Thus $n_1+\cdots+n_d=L$.
After the first bond of branch $i$ has been added, the remaining tree $T_i$, rooted at the other end of this bond, has $n_i-1$ bonds.
The growth orders of the $d$ branches can be interlaced in $L!/(n_1!\cdots n_d!)$ ways.
The induction hypothesis then gives
\eq
N(T)=\frac{L!}{\prod_{i=1}^d n_i!}
      \prod_{i=1}^d\frac{(n_i-1)!}{W(T_i)}
    =\frac{L!}{\prod_{i=1}^d n_iW(T_i)}
    =\frac{L!}{W(T)}.
\en
This proves the lemma.\hfill\qedm

\subsection{The extraction lemma}\label{S:lemma}
We shall prove an elementary lemma that allows us to extract a tree of a prescribed size without making its weight too large.
The lemma applies to any finite rooted tree, whether or not it is embedded in a lattice.

\begin{figure}[ht]
\centerline{\epsfig{file=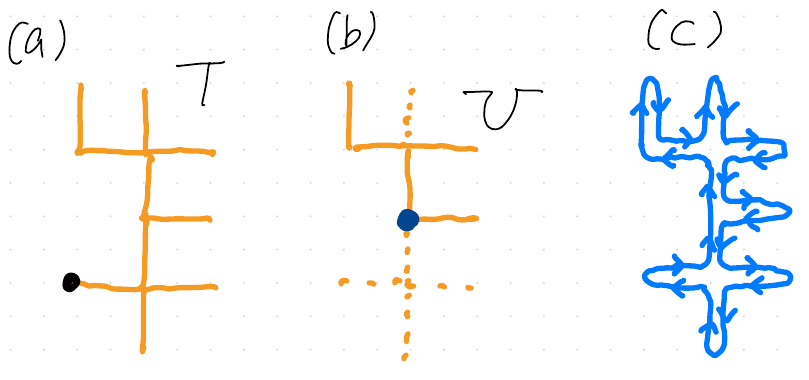,width=8truecm}}
\caption[dummy]{
(a)~The original rooted tree $T$.
(b)~A nonempty connected set of bonds $U\subset T$ (solid lines), rooted at its site closest, in tree distance, to the root of $T$.
(c)~A closed walk that traverses every bond of $T$ exactly twice, once in each direction.
}
\label{f:TU}
\end{figure}

We first explain how to choose the root of a subtree.
Let $U\subset T$ be a nonempty connected set of bonds.
There is a unique site of $U$ closest, in tree distance, to the root of $T$.
Since $U$ is connected and $T$ has no loops, the path from the root of $T$ to any site of $U$ passes through this site.
We choose it as the root of $U$, as in Figure~\ref{f:TU}(b).
Then every bond of $U$ has the same orientation as it has in $T$, and its downstream bonds in $U$ form a subset of its downstream bonds in $T$.
Consequently,
\eq
w_U(b)\le w_T(b)\quad\text{for any $b\in U$}.
\lb{monotonicity}
\en
It is important that $U$ need not contain the root of $T$.

\begin{lemma}[Extraction lemma]\label{L:extraction}
Let $T$ be a finite rooted tree with $M\ge1$ bonds.
For every integer $L$ with $1\le L\le M$, there exists a connected subtree $S\subset T$ with exactly $L$ bonds, rooted at its site closest to the root of $T$, such that
\eq
\log W(S)\le\frac{2L-1}{M}\log W(T).
\lb{extraction}
\en
\end{lemma}

\noindent
{\bf Proof.}
Assign to each bond $b\in T$ the nonnegative cost
\eq
 a(b)=\log w_T(b)\ge0.
\lb{cost}
\en
The total cost is $\sum_{b\in T}a(b)=\log W(T)$.

Choose a closed walk on $T$ that traverses every bond exactly twice, once in each direction.
See Figure~\ref{f:TU}(c).
Such a walk can be obtained by exploring each branch and returning along the bond by which it was entered.
Let
\eq
b_0,b_1,\ldots,b_{2M-1}
\en
be the successive bonds traversed by the walk, without their traversal directions.
We interpret the indices modulo $2M$, so that $b_{s+2M}=b_s$.
A segment may therefore continue from the end of this sequence back to its beginning.

Set $k=2L-1$.
For each starting position $s\in\{0,1,\ldots,2M-1\}$, let
\eq
A_s=\sum_{t=0}^{k-1}a(b_{s+t})
\lb{As}
\en
be the cost of the next $k$ steps.
Every step position of the closed walk belongs to exactly $k$ of these segments.
Since every bond occurs twice in the whole walk, we have
\eqa
\frac{1}{2M}\sum_{s=0}^{2M-1}A_s
 &=\frac{k}{2M}\sum_{s=0}^{2M-1}a(b_s)\nl
 &=\frac{k}{M}\sum_{b\in T}a(b)
  =\frac{2L-1}{M}\log W(T).
\lb{average}
\ena
There is therefore a starting position $s$ such that
\eq
A_s\le\frac{2L-1}{M}\log W(T).
\lb{selected}
\en

Let $U$ be the set of distinct bonds visited in this selected segment.
Because the segment is a walk, $U$ is connected, and hence is a subtree of $T$.
Furthermore, $k=2L-1<2M$, so no step position in the closed walk is used twice in the segment.
Every bond occurs only twice in the whole walk, and hence at most twice in the segment.
It follows that
\eq
|U|\ge\left\lceil\frac{2L-1}{2}\right\rceil=L,
\lb{usize}
\en
where $|U|$ denotes the number of bonds in $U$.

Root $U$ at its site closest to the root of $T$.
If $|U|>L$, repeatedly remove an endpoint (a site of degree one) distinct from this root, along with its incident bond, until exactly $L$ bonds remain.
This produces a connected subtree $S$ with the same root.
Using \rlb{monotonicity}, the nonnegativity of the costs, and the fact that every bond of $U$ is visited at least once in the selected segment, we find
\eqa
\log W(S)
 &=\sum_{b\in S}\log w_S(b)\nl
 &\le\sum_{b\in S}a(b)
  \le\sum_{b\in U}a(b)
  \le A_s
  \le\frac{2L-1}{M}\log W(T),
\ena
where we used $a(b)=\log w_T(b)\ge\log w_S(b)$ for $b\in S$.
This proves the lemma.\hfill\qedm

\medskip\noindent
{\em Remark:}\/ The selected walk segment is used only to find a connected set of bonds; it is not a growth history of the extracted tree.
In particular, the root of $S$ is not necessarily the first site of that segment.
The choice of root described above is what justifies \rlb{monotonicity}.

\subsection{Bouch's result and the proof of the main theorem}
The essential input is the following result of Bouch \cite{Bouch2015}, stated in terms of the tree weight.
We review its proof in Section~\ref{S:Bouch}.

\begin{proposition}[Bouch's theorem]\label{P:Bouch}
There are a constant $\CB>1$ and a sequence of rooted trees $T_1,T_2,\ldots$ on $\bbZ^2$ such that their bond numbers $L_j$ tend to infinity and
\eq
W(T_j)\le\CB^{L_j},\quad j=1,2,\ldots.
\lb{W<C}
\en
\end{proposition}

By Lemma~\ref{L:Kupi}, \rlb{W<C} is equivalent to $N(T_j)\ge(L_j)!/\CB^{L_j}$.

\medskip\noindent
{\bf Proof of Theorem~\ref{T:main}.}
Fix an arbitrary positive integer $L$.
Choose any tree $T_j$ in Proposition~\ref{P:Bouch} with $L_j\ge L$.
Lemma~\ref{L:extraction} gives a connected subtree $S_L\subset T_j$ with exactly $L$ bonds and
\eq
\log W(S_L)\le\frac{2L-1}{L_j}\log W(T_j)
             \le(2L-1)\log\CB.
\en
The square-lattice embedding is inherited from $T_j$.
The root of $S_L$ may differ from that of $T_j$, which is allowed in our setting.
A translation places the new root at the origin when desired.
Using Lemma~\ref{L:Kupi}, we obtain
\eq
N(S_L)\ge\frac{L!}{\CB^{2L-1}}
       \ge\frac{L!}{(\CB^2)^L}.
\lb{allL}
\en
Since $L$ was arbitrary, this proves the theorem with $C=\CB^2$.\hfill\qedm

The proof of the extraction lemma uses neither a bound on the vertex degrees nor any property of the square lattice.
The lattice enters only through Bouch's construction, which supplies arbitrarily large trees with the bound \rlb{W<C}.
Thus the additional argument does not replace Bouch's construction; it extracts the all-sizes conclusion from it.

\section{Bouch's construction}\label{S:Bouch}
We shall now prove Proposition~\ref{P:Bouch} by following Bouch's hierarchical construction.

\subsection{The basic structure of Bouch's trees}
Let $\ell_1,\ell_2,\ldots$ and $b_2,b_3,\ldots$ be sequences of positive integers satisfying the conditions below.
We shall specify the sequences explicitly later.

\begin{figure}[htbp]
\centerline{\epsfig{file=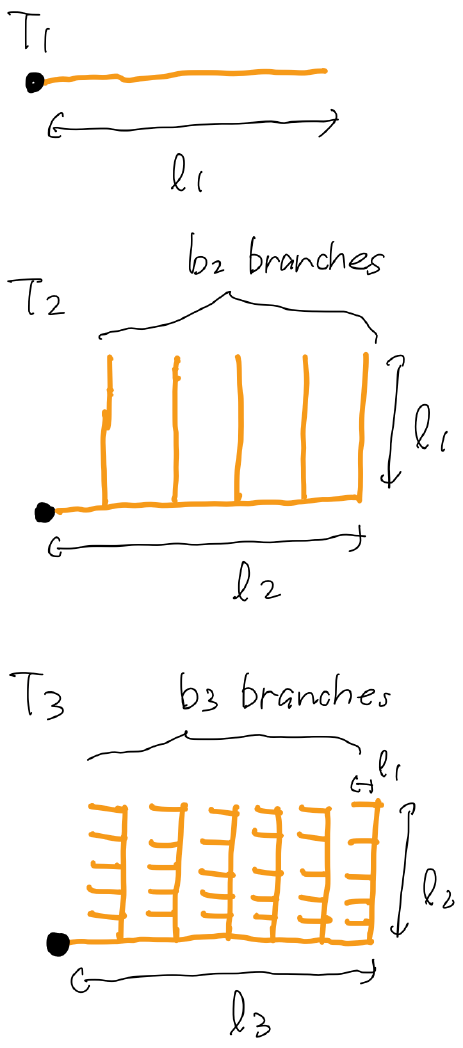,width=4.7truecm}}
\caption[dummy]{
Schematic figures of the first three generations, $T_1$, $T_2$, and $T_3$, of Bouch's sequence of trees.
}
\label{f:T123}
\end{figure}

Let us construct a sequence of rooted trees, $T_1,T_2,\ldots$, recursively as follows.
See Figure~\ref{f:T123}.
The tree $T_1$ is simply a horizontal segment consisting of $\ell_1$ bonds, where the left end of the segment is identified as the root.
The tree $T_2$ consists of a horizontal segment of $\ell_2$ bonds (whose left end is the root) from which $b_2$ vertical branches of length $\ell_1$ are sticking out.
The right-most branch is attached to the right end of the horizontal segment, and neighboring branches are separated by distance $\ell_2/b_2$.
The tree $T_3$ is constructed similarly, but now vertical branches are rotated copies of the tree $T_2$.

\begin{figure}[htbp]
\centerline{\epsfig{file=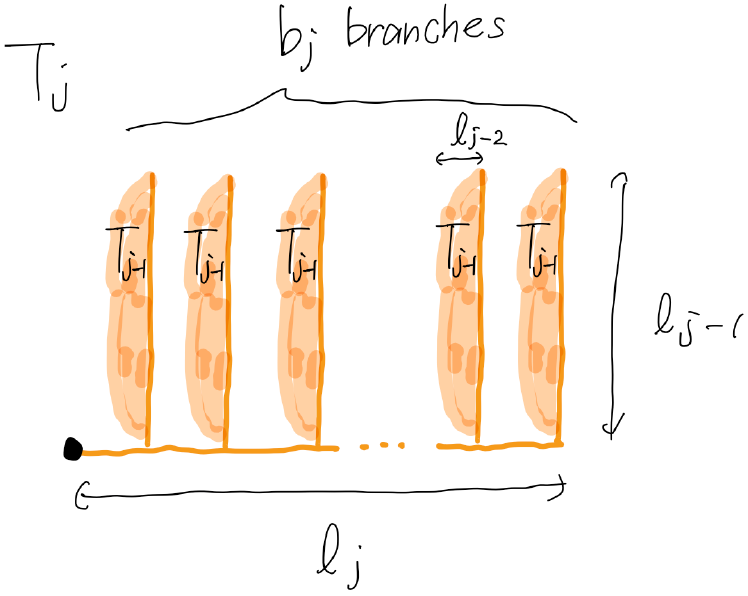,width=7truecm}}
\caption[dummy]{
The tree $T_j$ consists of a horizontal segment with $\ell_j$ bonds and $b_j$ vertical branches sticking out of it.
Each branch is a rotated copy of the tree $T_{j-1}$.
We shall choose various parameters so that most of the bonds in $T_j$ belong to branches with the youngest generations when $j$ is large.
}
\label{f:Tj}
\end{figure}

The procedure can be repeated and generalized.
As in Figure~\ref{f:Tj}, the tree $T_j$ is constructed by attaching $b_j$ copies of $T_{j-1}$, rotated counterclockwise by $90$ degrees, to a horizontal segment of $\ell_j$ bonds.
The attachment sites are at distances $\ell_j/b_j,2\ell_j/b_j,\ldots,\ell_j$ from its left end, which is the root.
It is convenient to require that $b_j$ divides $\ell_j$ for $j\ge2$.
We also need the condition
\eq
\frac{\ell_j}{b_j}>\ell_{j-2},
\lb{cond1}
\en
for $j\ge3$ to ensure that vertical branches do not overlap.
To see this more explicitly, place the root of each $T_j$ at the origin.
The tree $T_2$ lies in $[0,\ell_2]\times[0,\ell_1]$, with only its root on the vertical axis.
Inductively, if $T_{j-1}$ lies in $[0,\ell_{j-1}]\times[0,\ell_{j-2}]$ and has only its root on the vertical axis, its rotated copy at the $m$th attachment site lies in
\eq
\Bigl[\frac{m\ell_j}{b_j}-\ell_{j-2},\,
      \frac{m\ell_j}{b_j}\Bigr]\times[0,\ell_{j-1}].
\en
Condition \rlb{cond1} makes these strips disjoint and places them strictly to the right of the vertical axis.
Each copy meets the horizontal segment only at its root.
The resulting graph is therefore a tree in $[0,\ell_j]\times[0,\ell_{j-1}]$, again with only its root on the vertical axis.

Let $L_j$ denote the total number of bonds in the tree $T_j$.
From Figure~\ref{f:T123}, one sees
\eqg
L_1=\ell_1,\\
L_2=\ell_2+b_2\,\ell_1,\\
L_3=\ell_3+b_3L_2=\ell_3+b_3\,\ell_2+b_3b_2\,\ell_1,
\eng
which are generalized to
\eq
L_j=\ell_j+\sum_{k=1}^{j-1}b_jb_{j-1}\cdots b_{k+1}\,\ell_k.
\lb{Lj}
\en

\subsection{The recursion relation for the weights of the trees}
We shall derive the basic recursion relation for the weight $W_j:=W(T_j)$ of the tree $T_j$.
Since all the branches (which are copies of $T_{j-1}$) are ``downstream'' from the horizontal segment, each branch has a weight equal to $W_{j-1}$.
We thus have
\eq
W_j=(W_{j-1})^{b_j}\prod_{b\,\in\,{\rm hor. seg.}}w(b),
\en
where the product is over $\ell_j$ bonds on the horizontal segment.
Since we obviously have $w(b)\le L_j$ for any bond $b\in T_j$, we find\footnote{
Here, we do not faithfully follow Bouch, who writes down a tighter (and more complicated) upper bound.
It turns out that our rough bound is enough for the proof.
}
\eq
W_j\le(W_{j-1})^{b_j}(L_j)^{\ell_j},
\lb{Wrec}
\en
for $j\ge2$.
This is the basic recursion relation for the weight $W_j$.

\subsection{Description in terms of \texorpdfstring{$E_j$}{Ej}}
Let $E_0,E_1,\ldots$ be a strictly increasing sequence of positive integers such that $E_{j-1}$ divides $E_j$ for $j\ge2$.
We shall specify the sequence explicitly later.
We then write the sequences $\ell_1,\ell_2,\ldots$ and $b_2,b_3,\ldots$ in terms of $E_0,E_1,\ldots$ as
\eqg
\ell_1=E_1,\quad \ell_j=\frac{4E_jE_{j-2}}{E_{j-1}},\lb{lE}\\
b_j=\frac{E_j}{E_{j-1}},\lb{bE}
\eng
where $j\ge2$.
Then $b_j$ and $\ell_j$ are positive integers, and $b_j$ divides $\ell_j$, since $\ell_j=4E_{j-2}b_j$.
The non-overlap condition \rlb{cond1} is also satisfied.
Indeed, $\ell_3/b_3=4E_1>\ell_1$, while for $j\ge4$ we have
\eq
\frac{\ell_j}{b_j}=4E_{j-2}
 >\frac{4E_{j-2}E_{j-4}}{E_{j-3}}=\ell_{j-2},
\en
where we used $E_{j-4}<E_{j-3}$.

Let us rewrite $L_j$ of \rlb{Lj}, the total number of bonds in $T_j$, in terms of $E_0,E_1,\ldots$.
From \rlb{lE} and \rlb{bE}, we see
\eq
b_jb_{j-1}\ldots b_{k+1}\,\ell_k=\frac{E_j}{E_{j-1}}\frac{E_{j-1}}{E_{j-2}}\cdots\frac{E_{k+1}}{E_k}\,\frac{4E_kE_{k-2}}{E_{k-1}}
=\frac{4E_jE_{k-2}}{E_{k-1}},
\en
for $k\ge2$, and
\eq
b_jb_{j-1}\ldots b_{2}\,\ell_1=\frac{E_j}{E_{j-1}}\frac{E_{j-1}}{E_{j-2}}\cdots\frac{E_{2}}{E_1}\,E_1=E_j.
\lb{1st}
\en
It is worth noting that this is the total number of bonds in the first-generation branches in $T_j$.
Substituting these and \rlb{lE} into \rlb{Lj}, we get
\eq
L_j=\frac{4E_jE_{j-2}}{E_{j-1}}+\sum_{k=2}^{j-1}\frac{4E_jE_{k-2}}{E_{k-1}}+E_j
=E_j\Bigl\{1+4\sum_{k=2}^{j}\frac{E_{k-2}}{E_{k-1}}\Bigr\}.
\lb{Lj2}
\en

We next substitute \rlb{lE} and \rlb{bE} into the recursion relation \rlb{Wrec} to rewrite it as
\eq
W_j\le(W_{j-1})^{E_j/E_{j-1}}(L_j)^{4E_jE_{j-2}/E_{j-1}}.
\en
By taking the logarithm and dividing by $E_j$, we have
\eq
\frac{\log W_j}{E_j}\le\frac{\log W_{j-1}}{E_{j-1}}+\frac{4E_{j-2}}{E_{j-1}}\log L_j.
\en
Using this recursively, we finally get
\eq
\frac{\log W_j}{E_j}\le\frac{\log W_1}{E_1}+\sum_{k=2}^j\frac{4E_{k-2}}{E_{k-1}}\log L_k,
\lb{logW}
\en
for $j\ge2$.
This is our basic bound for the weight.

\subsection{Final estimates}
We shall now specify the sequence $E_0,E_1,\ldots$, thus specifying the sequences $\ell_1,\ell_2,\ldots$ and $b_2,b_3,\ldots$ as well.
Let $a_0$ be a positive integer.
Bouch sets $a_0=20$, but one may simply set $a_0=1$.
We then define a sequence $a_0,a_1,\ldots$ recursively by
\eq
a_j=2^{a_{j-1}}.
\lb{aj}
\en
We thus have, with Bouch's choice, $a_0=20$, $a_1=2^{20}$, $a_2=2^{2^{20}}$, $a_3=2^{2^{2^{20}}}$, and so on.
We then set
\eq
E_j=(a_j)^2.
\lb{Ej}
\en
For $j\ge2$, both $a_{j-1}$ and $a_j$ are powers of two, with $a_{j-1}<a_j$.
Hence $E_{j-1}$ divides $E_j$, as required above.

Noting that
\eq
\frac{E_{k-2}}{E_{k-1}}=\Bigl(\frac{a_{k-2}}{a_{k-1}}\Bigr)^2=\Bigl(\frac{a_{k-2}}{2^{a_{k-2}}}\Bigr)^2,
\en
the summation on the right-hand side of \rlb{Lj2} is bounded as
\eq
0\le4\sum_{k=2}^{j}\frac{E_{k-2}}{E_{k-1}}=4\sum_{k=2}^{j}\Bigl(\frac{a_{k-2}}{2^{a_{k-2}}}\Bigr)^2
\le4\sum_{k=2}^{\infty}\Bigl(\frac{a_{k-2}}{2^{a_{k-2}}}\Bigr)^2=:\epsilon_0<\infty,
\lb{conv}
\en
for any $j\ge2$.
We noted that the infinite sum converges (very rapidly) since a much larger sum $\sum_{a=a_0}^\infty(a/2^a)^2$ converges.
We then find from \rlb{Lj2} that
\eq
E_j\le L_j\le(1+\epsilon_0)E_j.
\lb{ELE}
\en
We note that, with the choice $a_0=20$, the positive constant $\epsilon_0$ is essentially determined by the first term $4(a_0/2^{a_0})^2$ in the sum. We have $\epsilon_0\simeq 1.5\times10^{-9}$.

The rapid convergence in \rlb{conv} reflects the fact that most of the bonds in $T_j$ are concentrated in the branches of the youngest generations.
With Bouch's choice $a_0=20$, almost all of them are already in the first-generation branches.
Let $L^{(1)}_j$ be the number of bonds in the tree $T_j$ that belong to the first-generation branches.
Since we have noted that $L^{(1)}_j=E_j$ in \rlb{1st}, we see from \rlb{ELE} that
\eq
\frac{1}{1+\epsilon_0}\le\frac{L^{(1)}_j}{L_j}\le1.
\en
Also note that
\eq
\frac{\ell_j}{L_j}\le \frac{\ell_j}{E_j}=\frac{4E_{j-2}}{E_{j-1}}=4\Bigl(\frac{a_{j-2}}{2^{a_{j-2}}}\Bigr)^2,
\en
where the right-hand side converges to zero rapidly as $j$ increases.
We have observed that, in the tree $T_j$ with sufficiently large $j$, the horizontal segment at the bottom (recall Figure~\ref{f:Tj}) contains a vanishingly small fraction of bonds.

We shall now analyze our bound \rlb{logW} for the weight of the tree to complete the proof.
We first note from \rlb{aj} and \rlb{Ej} that
\eq
\log E_k=2\log a_k=2\log 2^{a_{k-1}}=(2\log2)a_{k-1}.
\en
Then the summand on the right-hand side of \rlb{logW} is evaluated with the help of \rlb{ELE} as
\eqa
\frac{4E_{k-2}}{E_{k-1}}\log L_k&\le \frac{4E_{k-2}}{E_{k-1}}\log E_k+\frac{4E_{k-2}}{E_{k-1}}\log(1+\epsilon_0)
\nl
&=4\Bigl(\frac{a_{k-2}}{a_{k-1}}\Bigr)^2(2\log2)a_{k-1}+4\Bigl(\frac{a_{k-2}}{a_{k-1}}\Bigr)^2\log(1+\epsilon_0)
\nl
&=(8\log2)\frac{(a_{k-2})^2}{2^{a_{k-2}}}+\{4\log(1+\epsilon_0)\}\frac{(a_{k-2})^2}{4^{a_{k-2}}}.
\ena
As before, we readily find that the sum of the quantities on the right-hand side converges rapidly when taken from $k=2$ to $k=\infty$.
We thus have for any $j\ge2$ that
\eq
\sum_{k=2}^j\frac{4E_{k-2}}{E_{k-1}}\log L_k\le \sum_{k=2}^\infty\frac{4E_{k-2}}{E_{k-1}}\log L_k=:C_1<\infty.
\en
Letting $C_2:=C_1+\log W_1/E_1$, we finally find from \rlb{logW} that
\eq
\log W_j\le C_2\,E_j\le C_2\,L_j,
\en
where we used \rlb{ELE}.
With a constant $\CB:=e^{C_2}>1$, we have
\eq
W_j\le\CB^{L_j},
\en
which is nothing but the desired bound \rlb{W<C}.
The same bound holds for $j=1$, since $W_1=(\ell_1)!$ and $C_2\ge\log W_1/E_1$.
Finally, $L_j\ge E_j\to\infty$ by \rlb{ELE}.
This completes the proof of Proposition~\ref{P:Bouch}.\hfill\qedm
 
%%%%%%%%%%%%%%%%%%%%
\begin{samepage}

\end{samepage}


\begin{thebibliography}{10}

\bibitem{Bouch2015}
G. Bouch, 
{\em Complex-Time Singularity and Locality Estimates for Quantum Lattice Systems}\/,
J. Math. Phys. {\bf 56}, 123303 (2015).
\\\url{https://arxiv.org/abs/1011.1875}

\bibitem{ParkerCaoAvdoshkinScaffidiAltman2019}
D.E. Parker, X. Cao, A. Avdoshkin, T. Scaffidi, and E. Altman,
{\em A Universal Operator Growth Hypothesis}\/,
Phys. Rev. X {\bf 9}, 041017 (2019).
\\\url{https://journals.aps.org/prx/abstract/10.1103/PhysRevX.9.041017}

\bibitem{ShiraishiTasaki2024}
N. Shiraishi and H. Tasaki,
{\em The $S=\frac{1}{2}$ XY and XYZ models on the two- or higher-dimensional hypercubic lattice do not possess nontrivial local conserved quantities}\/,
Ann. Henri Poincar\'e (2026), doi:10.1007/s00023-026-01699-8.
\\\url{https://arxiv.org/abs/2412.18504}

\bibitem{TasakiVideo2025}
H. Tasaki,
{\em Signatures of quantum chaos in quantum spin systems: imaginary-time evolution and operator growth}\/,
(YouTube video, 2025).
\\\url{https://youtu.be/Jt_pC7DHflw}

 
 \end{thebibliography}
\end{document}